\documentclass[12pt]{article}

\usepackage{amsmath}
\usepackage{amssymb}
\usepackage{times}
\usepackage{graphicx}
\usepackage{placeins}
\usepackage{float}
\usepackage{color}

\usepackage[numbers,sort&compress]{natbib}
\usepackage[
colorlinks=true,
urlcolor=blue,
anchorcolor=blue,
citecolor=blue,
filecolor=blue,
linkcolor=blue,
menucolor=blue,
linktoc=page,
pdftitle={Entanglement-to-Stress Relation from an Effective Holographic Description of the Transverse-Field Ising Model},
pdfauthor={Dong-Hee Kim, Chanyong Park, and Jihye Yoo},
pdfsubject={Entanglement-to-stress reconstruction, infrared regularity, and microscopic TFIM test},
pdfkeywords={transverse-field Ising model, entanglement entropy, holographic reconstruction, infrared regularity, finite-size stress}
]{hyperref} 

\makeatletter
\long\def\@makecaption#1#2{%
  \vskip\abovecaptionskip
  \sbox\@tempboxa{#1.\ #2}%
  \ifdim \wd\@tempboxa>\hsize
    #1.\ #2\par
  \else
    \global\@minipagefalse
    \hb@xt@\hsize{\hfil\box\@tempboxa\hfil}%
  \fi
  \vskip\belowcaptionskip}
\makeatother

\def\ArxivWithoutColon arXiv:#1\relax{arXiv #1}
\newcommand{\Eprint}[2]{\href{#1}{\textcolor{black}{\ArxivWithoutColon#2\relax}}\ignorespaces}

\let\oldthebibliography\thebibliography
\let\endoldthebibliography\endthebibliography
\renewenvironment{thebibliography}[1]{%
  \oldthebibliography{#1}%
  \setlength{\itemsep}{0pt}%
  \setlength{\parsep}{0pt}%
  \setlength{\parskip}{0pt}%
  \small%
}{%
  \endoldthebibliography%
}

\begin{document}

\pagenumbering{roman}

\baselineskip 6mm
\renewcommand{\thefootnote}{\fnsymbol{footnote}}


\newcommand{\nc}{\newcommand}
\newcommand{\rnc}{\renewcommand}



\def\be{\begin{eqnarray}}
\def\ee{\end{eqnarray}}
\def\nn{\nonumber\\}


\def\la{\label}
\def\eq#1{(\textcolor{blue}{\ref*{#1}})}
\newcommand{\Eqref}[1]{Eq.~(\textcolor{blue}{\ref*{#1}})}
\newcommand{\Figref}[1]{Fig.~\textcolor{blue}{\ref*{#1}}}
\newcommand{\Figsubref}[2]{Fig.~\textcolor{blue}{\ref*{#1}(#2)}}


\def\a{\alpha}
\def\b{\beta}
\def\g{\gamma}
\def\G{\Gamma}
\def\d{\delta}
\def\D{\Delta}
\def\e{\epsilon}
\def\et{\eta}
\def\ph{\phi}
\def\Ph{\Phi}
\def\ps{\psi}
\def\Ps{\Psi}
\def\l{\lambda}
\def\L{\Lambda}
\def\m{\mu}
\def\n{\nu}
\def\th{\theta}
\def\Th{\Theta}
\def\r{\rho}
\def\s{\sigma}
\def\S{\Sigma}
\def\ta{\tau}
\def\o{\omega}
\def\O{\Omega}
\def\pr{\prime}


\def\half{\frac{1}{2}}

\def\goto{\rightarrow}

\def\na{\nabla}
\def\grad{\nabla}
\def\curl{\nabla\times}
\def\div{\nabla\cdot}
\def\pa{\partial}
\def\fr{\frac}

\def\bra{\left\langle}
\def\ket{\right\rangle}
\def\lb{\left[}
\def\lc{\left\{}
\def\ls{\left(}
\def\lp{\left.}
\def\rp{\right.}
\def\rb{\right]}
\def\rc{\right\}}
\def\rs{\right)}


\def\td#1{\tilde{#1}}


\def\Tr{{\rm Tr}\,}
\def\det{{\rm det}}
\def\inf{\infty}


\begin{titlepage}


\hspace{13cm}
 
\vspace{1cm}

\begin{center}
{\Large \bf Entanglement-to-Stress Relation from an Effective \\
Holographic Description of the Transverse-Field Ising Model}

\vskip 1cm
{Dong-Hee Kim\footnote{e-mail dongheekim@gist.ac.kr},
 Chanyong Park\footnote{e-mail cyong21@gist.ac.kr}, and
Jihye Yoo\footnote{e-mail jihye2288@gm.gist.ac.kr }}

\vskip 0.5cm

\textit{Department of Physics and Photon Science, Gwangju Institute of Science and Technology, 123 Cheomdangwagi-ro, Buk-gu, Gwangju 61005, Korea} \\

\end{center}

\thispagestyle{empty}

 \vskip 1cm


\centerline{\bf ABSTRACT} \vskip 4mm

We investigate an effective holographic description of the transverse-field Ising model (TFIM). We reconstruct the spatial metric from its entanglement entropy and determine the temporal metric using the Einstein-scalar equations, ultraviolet time normalization, and a smooth, horizonless center. Holographic renormalization then gives an enthalpy density that can be expressed through the entropy curvature at half the system size. This relation is a prediction of the effective gravity for the microscopic theory. We test it against the exact finite-size TFIM stress, which is not used in the reconstruction. Away from criticality, the gravity prediction agrees closely with the exact stress in the near-critical scaling regime. 
This result suggests an entanglement-to-stress relation in the massive Ising field theory, whose field-theoretic derivation remains open. We also find that the reconstructions for different system sizes and couplings share a common scalar potential, up to finite-size deviations near the center that move away as the size increases. This supports the view that the reconstructed geometries are solutions of a single effective Einstein-scalar theory.

\vspace{1cm}


\end{titlepage}

\pagenumbering{arabic}
\renewcommand{\thefootnote}{\arabic{footnote}}
\setcounter{footnote}{0}

\makeatletter
\renewcommand*\l@section[2]{%
  \ifnum \c@tocdepth >\z@
    \addpenalty\@secpenalty
    \setlength\@tempdima{1.5em}%
    \begingroup
      \parindent \z@ \rightskip \@pnumwidth
      \parfillskip -\@pnumwidth
      \leavevmode \bfseries
      \advance\leftskip\@tempdima
      \hskip -\leftskip
      #1\nobreak\hfil
      \nobreak\hb@xt@\@pnumwidth{\hss #2}\par
    \endgroup
  \fi}
\makeatother
\tableofcontents


\section{Introduction}

The AdS/CFT correspondence provides a useful framework for studying strongly interacting systems in QCD and condensed matter physics \cite{Maldacena:1997re,Gubser:1998bc,Witten:1998qj,Witten:1998zw}. It gives a simple geometric description for many nontrivial physical observables of strongly interacting quantum field theories. In a standard bottom-up construction, one chooses a bulk theory and computes boundary observables, while the microscopic Hamiltonian behind the geometry remains unknown. In the present work, we go in the opposite direction. We start from a known lattice Hamiltonian and ask whether its entanglement structure can determine an effective bulk geometry that predicts other observables.

Since the radial direction of the bulk is related to the renormalization group (RG) scale \cite{deBoer:1999tgo,Heemskerk:2010hk,Faulkner:2010jy,Kim:2016ayz}, reconstruction of the dual gravity requires data over an entire range of length scales. Entanglement entropy is natural for this purpose, because intervals of different sizes probe different scales and the Ryu-Takayanagi prescription relates their entropies to the bulk geometry \cite{Ryu:2006bv,Ryu:2006ef}. Metric reconstruction has been developed with Abel inversion, differential entropy, kinematic space, relative entropy, and related methods \cite{Bilson:2008ab,Bilson:2010ff,Czech:2014wka,Czech:2015qta,Blanco:2013joa,Lashkari:2013koa,Jafferis:2015del,Hamilton:2005ju,Swingle:2009bg,Nozaki:2012zj,Saha:2018BGE}. In particular, Ref.~\cite{Cao:2020uvb} reconstructed a bulk metric from the entanglement entropy of a one-dimensional free-fermion chain by using the tensor Radon transform. Recently, holographic reconstruction methods based on perturbative minimal surfaces and machine learning have also been explored \cite{Jokela:2025ime,Park:2022fqy,Park:2023slm,Ahn:2024jkk}. The bulk metric has also been reconstructed from noisy entanglement data and from lattice Yang-Mills entanglement \cite{Jokela:2020TPH,Jokela:2023DGD}. Since equal-time entanglement fixes only the spatial metric, the temporal metric has been reconstructed with an additional observable such as Wilson loops \cite{Filev:2026gka}, with unequal-time entanglement \cite{Chae:2026wnp} or timelike entanglement \cite{Feng:2026MRT,AliAkbari:2026HEG}, or by completing the Abel-inverted geometry into an effective gravity theory with a metric ansatz for the temporal component \cite{Huh:2026nvt}.

In the present work, we extend the method of Ref.~\cite{Huh:2026nvt} to a system on a circle and apply it to the one-dimensional transverse-field Ising model (TFIM). The TFIM is exactly solvable, and near its quantum critical point it is described by the Ising conformal field theory (CFT) deformed by the energy operator \cite{Pfeuty:1970qrn,Sachdev:2011fcc,Mbeng:2020awt}. In the disordered phase, its scaling limit is a free massive Majorana theory. Therefore, we can calculate the entanglement entropy and the finite-size stress independently and exactly. Related microscopic and holographic studies of the TFIM can be found in Refs.~\cite{Kim:2016hig,Kim:2020kma,Kim:2020nst,Bamba:2023ohp}. After fixing the spatial metric from the entanglement entropy, we fix the temporal metric with an Einstein-scalar model and regularity at the center. Since the energy operator has dimension one, the dual scalar lies at the Breitenlohner-Freedman (BF) bound, which fixes the logarithmic ultraviolet structure (Appendix~\ref{Appendix: BFentropy}).

For a smooth, concave single-interval entropy with suitable asymptotic behavior \cite{Casini:2004bw}, Abel inversion reconstructs a spatial metric, and the forward geodesic calculation returns the input entropy by construction. A nontrivial test therefore uses an observable that does not enter the reconstruction. Here this observable is the finite-size stress obtained from the exact TFIM spectrum. The construction has three steps. Entropy determines the spatial geometry, the Einstein-scalar equations and boundary conditions select its temporal completion, and holographic renormalization predicts the stress. The off-critical comparison tests this prediction against the microscopic theory. At criticality, the exact Ising entropy reconstructs global AdS$_3$ analytically, and the Casimir energy checks the sign and normalization. Away from criticality, the relevant deformation generates an RG flow \cite{Park:2024fik,Park:2021nyc}, and the geometry is not known in advance.

Applying holographic renormalization to the reconstructed geometry, we calculate the boundary stress tensor. Let $f_{S_E}(r)$ be the radial metric function reconstructed from the entanglement entropy. Then the holographic enthalpy density is
\be
(\e+p)_{\rm grav}(S_E)
=-\fr{1}{8\pi GR^2{\cal H}_\infty(f_{S_E})}
\quad {\rm with} \quad
{\cal H}_\infty(f)=\lim_{r\to\infty}\int_0^r ds\,\fr{1-f(s)^{-1}}{s^2},
\la{Result: entanglementToStress}
\ee
where $G$ is the three-dimensional Newton constant and $R$ is the AdS radius. This is the quantity that we compare with the exact TFIM stress. Using the geodesic turning-point relation, we can also express it entirely in terms of the entanglement entropy
\be
(\e+p)_{\rm grav}=\fr1\pi\left.\fr{d^2S_E}{d\ell^2}\right|_{\ell=L/2} .
\la{Result: introRelation}
\ee
This relation holds for the regular geometries specified in Sec.~\ref{Section: BrownYork}, for which the spatial entropy data determine the holographic enthalpy density. At criticality, the relation also follows from the Calabrese--Cardy entropy and the vacuum Casimir energy \cite{Calabrese:2004eu,Bloete:1986qm,Affleck:1986bv}, and from saturation of the quantum null energy condition \cite{Bousso:2015mna,Ecker:2019ocp}. Away from criticality, it is a prediction of the effective holographic construction, while, to our knowledge, its direct field-theory derivation remains open. 
The stress predicted from the reconstructed geometry agrees closely with the independently calculated exact TFIM stress in the  near-critical scaling regime. The comparison as a function of the coupling and the collapse against a dimensionless scaling variable test the predictive content of the construction beyond reproducing its input entropy. 

Ref.~\cite{Huh:2026nvt} introduced the completion of the Abel-inverted geometry into an effective gravity theory. In that work, the boundary is a line, and a metric ansatz such as a Lorentz-invariant domain wall fixes the temporal metric. On a circle, the finite size breaks the boundary Lorentz invariance and produces a nonzero stress, so this ansatz cannot be used. Instead, we determine the temporal metric by the bulk equations and center regularity, and its normalizable part gives the finite-size stress. Compared with Ref.~\cite{Huh:2026nvt}, this work adds four points. First, we derive the holographic prediction in \Eqref{Result: introRelation}, which is not yet known in field theory away from criticality. Second, we identify the conditions on which it relies. Third, we test it directly against the exact stress of a microscopic Hamiltonian, including the continuum limit. Fourth, we show that one universal scalar potential describes the reconstructions for different sizes and couplings in the scaling regime, so that the reconstructed gravity behaves as one effective theory.

The rest of this paper is organized as follows. In Sec.~\ref{Section: TFIM}, we summarize the exact spectrum and entanglement entropy of the TFIM. In Sec.~\ref{Section: critical}, the critical reconstruction establishes the analytic test and stress normalization. Sec.~\ref{Section: Relevant deformation} develops the off-critical reconstruction and culminates in the independent stress comparison. Sec.~\ref{Section: universalPotential} shows that one universal scalar potential describes the reconstructions for different sizes and couplings. Sec.~\ref{Section: discussion} discusses its interpretation and limitations.

\section{The TFIM as an exactly solvable test system}\la{Section: TFIM}

The one-dimensional TFIM is exactly solvable \cite{Lieb:1961fr,Pfeuty:1970qrn,Suzuki:2013,Sachdev:2011fcc,Mbeng:2020awt} and provides a numerically computable entanglement entropy, which can be used as a probe of quantum criticality
\cite{Latorre:2003kg,Jin:2003lzd,Calabrese:2004eu,Calabrese:2009qy,Suzuki:2013,Sachdev:2011fcc}.
The Hamiltonian is
\be
H = - J \sum_{i=1}^N \s^x_i  \, \s^x_{i+1} -  h \sum _{i=1}^N \s^z_i ,
\ee
where $J$ is the nearest-neighbor Ising coupling and $h$ is the transverse field. Here $i$ labels the lattice sites, and $\s_i^x$ and $\s_i^z$ are Pauli operators at site $i$. We denote the number of sites by $N$ and use periodic boundary conditions. Near its quantum critical point, the lattice Hamiltonian at long distances is described by the Ising CFT with a relevant deformation \cite{Pfeuty:1970qrn,Sachdev:2011fcc,Calabrese:2004eu}
\be
H &=& H_{\rm crit} + H_{\rm def} ,
\ee
where the critical lattice Hamiltonian at $J=h$ is given by
\be
H_{\rm crit}
=-J\left(\sum_i\s_i^x\s_{i+1}^x+\sum_i\s_i^z\right).
\ee
Its scaling limit is the Ising CFT, while the lattice spacing remains the microscopic regulator. A $Z_2$-even perturbation describes the deviation from criticality
\be
H_{\rm def}=-(h-J)\sum_i\s_i^z .
\ee

To connect to the continuum theory, we first take into account the quasiparticle spectrum. After the Jordan-Wigner transformation, the spin model becomes a fermionic system. Then, we can diagonalize it by applying the Bogoliubov transformation   \cite{Lieb:1961fr,Pfeuty:1970qrn,Suzuki:2013,Sachdev:2011fcc,Mbeng:2020awt}. We summarize the parity convention of the finite chain and the prescription for the ground-state energy in Appendix~\ref{Appendix: TFIMpressure}. The single-particle dispersion is
\be
\o_k=2\sqrt{(h-J\cos\bar k)^2+J^2\sin^2\bar k},
\la{Result: exactTFIMdispersion}
\ee
where $\bar k=k\bar a$ is a dimensionless lattice momentum and $\bar a$ is the lattice spacing. For small $k$, the dispersion becomes
\be
\o_k &\approx&  2  \sqrt{ | h - J |^2 +  h J  \bar{k}^2   }     = \sqrt{m_\ps^2 + v^2 k^2
}  ,   \la{Result: TFIMDR}
\ee
where $m_\ps =2|h-J|$ is the spectral gap and $v=2\sqrt{hJ}\,\bar a$ is the velocity. On the disordered branch with $h>J$, the continuum Hamiltonian becomes that of a free massive Majorana fermion \cite{Sachdev:2011fcc,Mbeng:2020awt},
\be
H_{\rm cont}=\int dx\,\left(i v\,\bar{\ps}\g^1\pa_x\ps
+m_\ps\bar{\ps}\ps\right). \la{Relation: fermionPic}
\ee
On a finite periodic chain, momentum quantization introduces the scale $v/L$. The gap approaches $m_\ps$ in the thermodynamic limit.

We divide the system into a subsystem $A$ of length $\ell$ and its complement of length $L-\ell$. For the free-fermion ground state, we obtain the entanglement entropy exactly from the restricted Majorana covariance matrix \cite{Peschel:2002yqj,Eisler:2009vye}. With its positive eigenvalues $\n_j$, the von Neumann entropy is
\be
S_E =  - \sum_{j=1}^{n} \left[  \fr{1+\n_j}{2} \log \left( \fr{1+\n_j}{2} \right) + \fr{1-\n_j}{2} \log \left( \fr{1-\n_j}{2} \right) \right] ,   \la{Relation: GFormu}
\ee
where $(1\pm\n_j)/2$ are the single-particle entanglement eigenvalues and $n=\ell/\bar a$ is the number of sites in $A$. For off-critical states with $h \ne J$, we evaluate the entanglement entropy with \Eqref{Relation: GFormu}. \Figref{Figure: TFIMentropy} plots the TFIM entanglement entropy with various $h$ at $L=512$. For $h>J$, the $\s_i^z$ deformation leads to the disordered phase, where $\langle\s_i^x\rangle=0$. Near criticality, the entropy of a large subsystem in the thermodynamic limit saturates as \cite{Calabrese:2004eu,Casini:2005rm}
\be
S_E = \fr{c}{3} \ \log  \ls \fr{\xi}{\bar{a}} \rs  + \widetilde S_0    ,
\ee
where $\widetilde S_0$ is nonuniversal and the scaling correlation length is $\xi\simeq v/m_\ps$. At the critical point with $h=J$, the correlation length diverges, and conformal symmetry is restored. Far into the disordered phase, the spins are polarized along the transverse field, so the ground state approaches a product state with vanishing entanglement.

\begin{figure}[t]
	\begin{center}
		\vspace{-0.5cm}
		\hspace{-0.5cm}
		\includegraphics[angle=0,width=0.5 \textwidth]{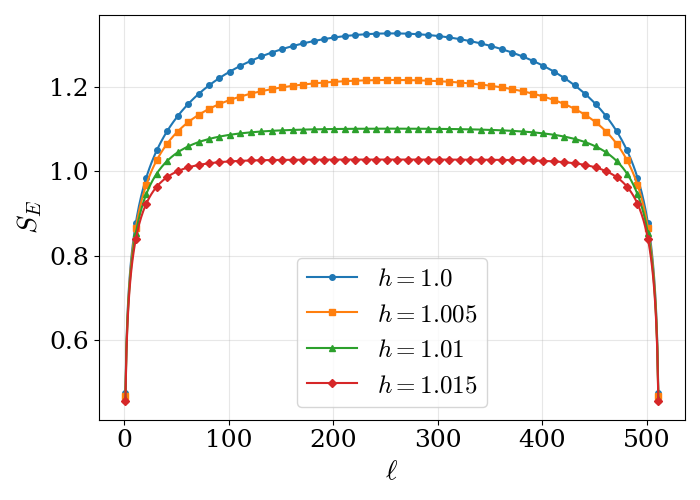}
		\vspace{-0cm}
			\caption{Periodic-TFIM ground-state entropy at $L=N=512$, $J=1$, and $\bar a=1$.}
			\la{Figure: TFIMentropy}
	\end{center}
\end{figure}

For the TFIM, we can evaluate the exact finite-size Casimir energy from the same quasiparticle spectrum \cite{Pfeuty:1970qrn,Mbeng:2020awt}. On the disordered branch, the physical ground state is the Neveu-Schwarz quasiparticle vacuum. After subtracting the extensive thermodynamic energy, we define the Casimir energy as $E_{\rm Cas}=E_0^{\rm NS}-N e_\inf$. Normalizing the energy density and pressure as
\be
\e= \frac{E_{\rm Cas} }{v L} \quad {\rm and} \quad
p=- \frac{1}{v} \,  \lp \frac{\pa E_{\rm Cas}}{\pa L}  \right|_{h,J,\bar a}  ,
\ee
the enthalpy density of the TFIM is given by $\e+p$, which is renormalization scheme-independent (see the details in Appendix~\ref{Appendix: TFIMpressure}).

Before closing this section, it is worth noting that we should use the continuum description only inside the Ising scaling regime. When the bare mass is small, the RG flow stays close to the Ising fixed point over a wide range of length scales. In that regime, the irrelevant lattice corrections become small before the mass deformation controls the infrared physics. As the mass increases, this scaling window becomes narrower, and the continuum description gradually loses accuracy. Below we reconstruct the dual gravity of the TFIM by using the entanglement entropy. In this case, we have to regard the reconstructed Einstein-scalar geometry as an effective description of the flow from the Ising CFT toward the massive infrared regime. It is not a geometrization of the full microscopic TFIM RG trajectory \cite{Wilson:1973jj,Polchinski:1983gv,Wegner:1972my,Fisher:1998kv,Kadanoff:1966wm,Cardy:1984epx}.

\section{Critical theory and analytic reconstruction}\la{Section: critical}

We begin at the critical point, where the required boundary quantities are known analytically. At the fixed point, there is no nontrivial RG flow, and the dual geometry is expected to be AdS. Therefore, this case gives a simple analytic check of the method before we turn to off-critical lattice data. At the critical point $J=h$, the long-distance theory becomes the Ising CFT, and conformal invariance gives \cite{Holzhey:1994we,Calabrese:2004eu,Latorre:2003kg,Jin:2003lzd,Casini:2009sr}
\be
S_E = \fr{c}{3}  \log \ls \fr{L}{\pi a} \ \sin \fr{\pi \ell}{L}  \rs + S_0  ,   \la{Result: unSEE}
\ee
where $c=1/2$ is the Ising central charge, $a$ is the continuum ultraviolet regulator, and $S_0$ is a nonuniversal additive constant. We identify the periodic chain of circumference $L$ with the boundary circle and use
\be
L=N\bar a=2\pi R  \quad {\rm and} \quad \ell=n\bar a .
\la{Relation: physicalcircumference}
\ee
This relation connects the chain length to the AdS radius $R$ of the reconstructed geometry. Then, the entropy becomes
\be
S_E = \fr{1}{6}  \log \ls \fr{  \g N}{\pi} \ \sin \fr{\pi n}{N} \rs  ,  \la{Result: CFTEn}
\ee
where $\g$ is a dimensionless matching factor defined by $S_0 = c/3 \log ( a\g/\bar{a} )$. For $\ell=L/2$ at criticality, the entropy grows logarithmically as $L\to\inf$.

On the lattice, $\s_i^z$ generates the transverse-field perturbation. In the continuum expansion, $\s_i^z$ contains an identity part, which shifts the vacuum energy. It also contains an energy-density part proportional to the Ising operator ${\cal O}_{\rm E}\sim\bar\ps\ps$ of dimension one. After subtracting the identity part, the connected critical correlator behaves as
\be
\bra \s^z_i \ \s^z_j \ket_{\rm conn}
\sim \fr{1}{|i-j|^2}  .
\ee
This agrees with the continuum Majorana correlator
\be
\bra  \bar{\ps}  \ps (x)  \ \bar{\ps} \ps (y)   \ket \sim \fr{1}{|x-y|^2}  .  \la{Result: TFIMcd}
\ee
At the same critical point, the finite-size ground-state energy has the
universal Casimir form \cite{Bloete:1986qm,Affleck:1986bv}
\be
E_{\rm Cas}=-\fr{\pi c v}{6L}+{\cal O}(L^{-3}),
\ee
so that $\e=p=-\pi c/(6L^2)$. 
The entanglement entropy in \Eqref{Result: unSEE} results in the following relation at half the system size
\be
\left.\fr{d^2S_E}{d\ell^2}\right|_{\ell=L/2}=-\fr{c\pi^2}{3L^2} .
\ee
Comparing it with the enthalpy density, we finally find that the critical theory satisfies
\be
\e+p=\fr1\pi\left.\fr{d^2S_E}{d\ell^2}\right|_{\ell=L/2} ,
\la{Result: criticalCurvatureStress}
\ee
which follows directly from known CFT results \cite{Calabrese:2004eu,Calabrese:2009qy,Bloete:1986qm,Affleck:1986bv}.

We can also understand \Eqref{Result: criticalCurvatureStress}  from the quantum null energy condition (QNEC) \cite{Ecker:2019ocp,Ecker:2020gnw}. We move the right endpoint of the interval along the null directions $x^\pm=t\pm x$. In two dimensions, QNEC states that
\be
2\pi\langle T_{\pm\pm}\rangle
\geq
\partial_\pm^2S_E+\frac{6}{c}(\partial_\pm S_E)^2.
\ee
For the CFT vacuum, the entropy separates into left- and right-moving contributions, $S_E=S_+(x^+)+S_-(x^-)$, and QNEC is saturated in both null directions \cite{Ecker:2019ocp,Koeller:2015qmn}. At $\ell=L/2$, one has $\partial_+S_E=\partial_-S_E=0$, while the factorized form implies $\partial_+\partial_-S_E=0$. At the critical point, therefore, the second derivative becomes
\be
\frac{d^2S_E}{d\ell^2}
=
\partial_+^2S_E+\partial_-^2S_E.
\ee
Adding the two saturated QNEC relations and using $\langle T_{++}+T_{--}\rangle=(\epsilon+p)/2$, we recover \Eqref{Result: criticalCurvatureStress}. Away from criticality, the mass term couples the left- and right-moving modes. The entropy no longer splits into two parts, so QNEC need not be saturated.  Consequently, QNEC alone does not guarantee that \Eqref{Result: criticalCurvatureStress} holds away from criticality. In this work, we will derive a conditional prediction of the same form even in the off-critical case by using the effective holographic construction, and test it independently against the off-critical TFIM stress.

We next ask what geometry this critical entanglement entropy gives. Since the entropy is known analytically on the circle at the critical point, we can also do the reconstruction analytically. 
We use the exact entropy as the boundary observable that determines the effective bulk geometry. The critical theory then fixes the overall gravitational normalization. In this picture, the asymptotic AdS region corresponds to the ultraviolet end of the continuum Ising scaling theory, not to the microscopic lattice cutoff. For the critical Ising CFT, the natural reference geometry is global AdS$_3$
\be
 ds^2=-\left(1+\fr{r^2}{R^2}\right)dt^2+r^2d\th^2
 +\fr{dr^2}{1+r^2/R^2}.
\la{Result: globalAdSmetric}
\ee
We now ask whether we can recover this geometry directly from the exact critical entropy, without imposing it from the outset. Since an equal-time geodesic probes only the spatial geometry, we keep its radial function unknown and write
\be
 ds_{\rm spatial}^2=r^2d\th^2+\fr{dr^2}{f(r)^2}.
\la{Ansatz: criticalSpatialMetric}
\ee
The critical reconstruction thus reduces to finding the single unknown function $f(r)$ from the entanglement entropy.

For \Eqref{Ansatz: criticalSpatialMetric}, the equal-time holographic entropy is the length of the bulk geodesic anchored on the same interval \cite{Ryu:2006bv,Ryu:2006ef,Hubeny:2007xt,Lewkowycz:2013nqa},
\be
S_E = \fr{1}{4 G} \int_{-\D\th/2}^{\D\th/2}
d\th \sqrt{r^2 +  \fr{1}{f^2}
\ls \fr{d r}{d \th}\rs^2 } .   \la{Equation: HoloEn}
\ee
Here $\D\th$ is the angular size of the boundary interval.  The radial coordinate reaches its minimum value $r_t$, corresponding to the turning point of the geodesic. Rotational symmetry expresses both $\D\th$ and the entropy in terms of $r_t$
\be
\D\th(r_t)
&=& \int_{r_t}^{\inf}dr\,\fr{2r_t}{rf\sqrt{r^2-r_t^2}},
\la{Relation: turningAngle}\\
S_E  (r_t)  &=& \fr{c}{3 R }\int_{r_t}^{\inf} dr \ \fr{ r }{  f  \sqrt{r^2 - r_t^2} } ,
\la{Relation: turningEntropy}
\ee
where $c=3R/(2G)$ is the Brown-Henneaux central charge. Matching it to the Ising central charge gives $3R/(2G)=1/2$. For an interval of $n$ sites, we have $\D\th=2\pi n/N$. These integrals satisfy the Hamilton-Jacobi relation
\be
\fr{d S_E (n)}{d n} = \fr{\pi \, r_t(n) }{2 G N}  .  \la{Equation: HamiltonJacobi}
\ee
This implies that we can find the subsystem size as a function of the turning point, $n(r_t)$, from the known $S_E(n)$. With this relation, the inverse Abel transformation determines the unknown function $f$ as
\be
\fr{1}{f} = - \fr{2 r^2}{N}   \int_r^{\inf} dr_t \ \fr{1}{\sqrt{r_t^2 - r^2}} \  \fr{d}{d r_t} \ls \fr{n(r_t)}{r_t}\rs . \la{Relation: IAbel}
\ee

The Hamilton-Jacobi relation in \Eqref{Equation: HamiltonJacobi} and the critical entanglement entropy in \Eqref{Result: CFTEn} give the subsystem size as a function of the turning point,
\be
n (r_t) = \fr{N}{\pi} {\rm arccot} \ls \fr{r_t}{R}\rs  .
\ee
Note that the renormalization factor $\g$ does not change this result. Substituting this function into the inverse Abel transformation in \Eqref{Relation: IAbel}, we finally obtain
\be
f (r) = \sqrt{ \fr{r^2}{R^2}  + 1 }.   \la{Result: AdSmetric}
\ee
The recovered radial function is exactly the spatial factor of the global AdS metric in \Eqref{Result: globalAdSmetric}. Thus the critical entanglement entropy reconstructs the expected global AdS spatial geometry, without imposing this radial function beforehand. The same geometry was also obtained from the CFT entropy on a circle in Ref.~\cite{Saha:2018BGE}.

We also carry out the critical reconstruction numerically with the finite-chain entropy, as shown in \Figref{Figure: criticalReconstruction}. The reconstructed spatial metric approaches the analytic result in \Eqref{Result: AdSmetric}. The forward geodesic calculation reproduces the CFT entanglement entropy again, as shown in \Figsubref{Figure: criticalReconstruction}{b}. The circumference fixes $R=L/(2\pi)$, while the additive entropy constant changes only the finite radial matching scale. In this numerical reconstruction, we preserve the exact symmetry $S_E(n)=S_E(N-n)$ and include the half-chain point $n=N/2$, where the turning point reaches the center. Near this point, we obtain the Hamilton-Jacobi derivative from an even polynomial fit to $S_E(n)$ in $n-N/2$. We then insert the resulting turning-point data directly into the inverse Abel transformation, with an asymptotically AdS continuation beyond the largest resolved turning point. We use the same numerical prescription for the off-critical data in the next section.

\begin{figure}[t]
	\begin{center}
		\vspace{-0.5cm}
		\hspace{-0.5cm}
		\begin{minipage}[t]{0.45\textwidth}
		\centering
		\includegraphics[angle=0,width=\textwidth]{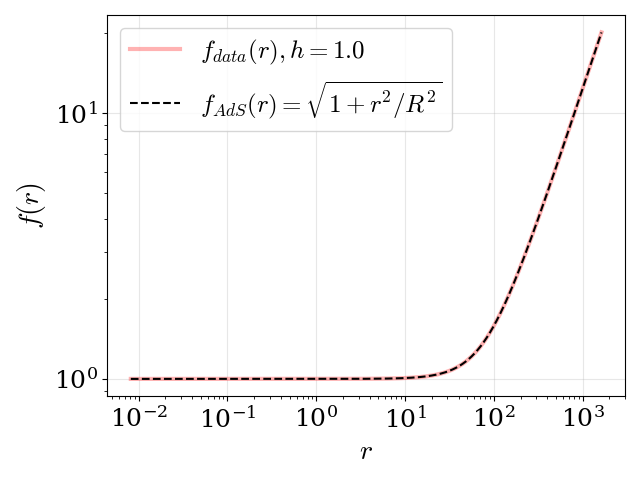}\\[-1mm]
		(a) $f(r)$
		\end{minipage}
		\hspace{0.5cm}
		\begin{minipage}[t]{0.45\textwidth}
		\centering
		\includegraphics[angle=0,width=\textwidth]{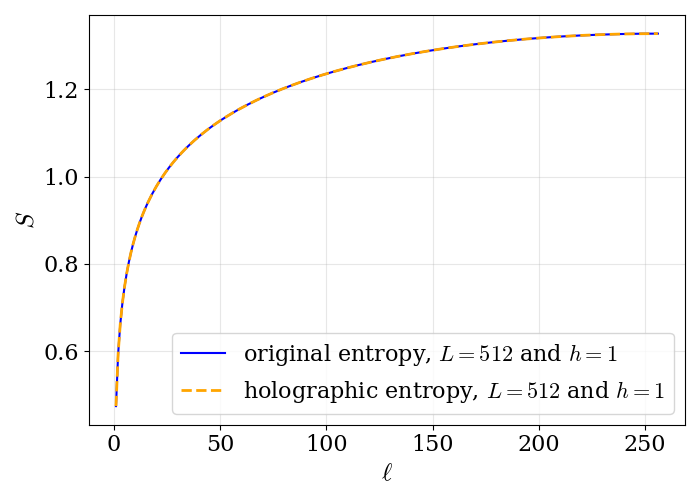}\\[-1mm]
		(b) Entropy from the reconstructed geometry
		\end{minipage}
		\vspace{-0cm}
			\caption{Critical reconstruction. (a) Numerically reconstructed spatial metric compared with the analytic global-AdS result. (b) Forward-geodesic entropy compared with the critical Ising-CFT entropy.}
			\la{Figure: criticalReconstruction}
	\end{center}
\end{figure}

\section{Relevant deformation, the entanglement-to-stress test, and the universal potential}  \la{Section: Relevant deformation}

We now turn to the central test of whether the off-critical geometry reconstructed from entanglement can provide independent predictions. In general, the relevant deformation generates a nontrivial RG flow \cite{Park:2024fik,Park:2021nyc}, so the effective bulk geometry is not known in advance. We first reconstruct its spatial metric from the TFIM entropy and determine its temporal metric by infrared regularity. We then extract the holographic stress and compare it with the one derived from the exact single-particle spectrum \cite{Pfeuty:1970qrn,Mbeng:2020awt}, without using that stress as reconstruction input.

\subsection{Reconstruction of effective gravity for the off-critical TFIM}\la{Section: fullGeometry}

Near the Ising critical point, a single scalar operator corresponding to the energy density generates the deformation. We therefore model the active sector minimally by the metric and one bulk scalar. From now on, we focus on local two-derivative scalar-tensor gravity. In this class, a field redefinition brings a nondegenerate one-scalar theory to the Einstein frame. After normalizing the scalar canonically, the remaining model dependence of this truncation is in the scalar potential. Higher-curvature and other higher-derivative terms may give corrections, but we do not include them in this leading-order description. We also require the reconstructed solution to stay on a regular branch without a curvature singularity. With these assumptions, the minimal gravity action is expressed as
\be
 S=\fr{1}{16 \pi G}\int d^3x\sqrt{-G}
 \left[
 {\cal R}-2\L-\fr12(\pa\ph)^2-\fr{V(\ph)}{R^2}
 \right]\quad {\rm with}\quad \L=-\fr1{R^2},
\la{Ansatz: Action}
\ee
where $R$ is the asymptotic AdS radius. For the static rotationally invariant states considered here, we use the following metric ansatz in the areal gauge
\be
 ds^2=-g(r)^2dt^2+r^2d\th^2+\fr{dr^2}{f(r)^2}.
\la{Ansatz: metric}
\ee
Unlike a standard bottom-up construction \cite{Freedman:1999gp,Gursoy:2007er,Park:2019pzo}, we do not start by choosing the potential $V(\ph)$. Instead, the entanglement entropy first determines the spatial function $f(r)$. Then, the bulk equations, the boundary normalization of time, and the regularity condition at the center fix the temporal function $g(r)$. Once we know $f$ and $g$, the equations also determine the scalar profile and the potential along the reconstructed trajectory \cite{Myers:2012ed}. For the ansatz above, the independent Einstein equations are given by
\be
 0&=&g''+\fr{f'}f\left(g'-\fr gr\right),
 \la{Equation: g}\\
 \ph'^2&=&\fr2r\left(\fr{g'}g-\fr{f'}f\right),
 \la{Equation: phi}\\
 V(r)&=&2-\fr{R^2f^2}{r}
 \left(\fr{f'}f+\fr{g'}g\right).
 \la{Equation: V}
\ee
Note that \Eqref{Equation: g} contains neither $\ph$ nor $V$. It follows from $T^t{}_t=T^\th{}_\th$, which holds for any static scalar profile $\ph(r)$.

The operator content of the Ising theory fixes the ultraviolet behavior of this model. The bulk scalar represents the energy-density deformation ${\cal O}_{\rm E}\sim\bar\ps\ps$, which has dimension $\D=1$. With the mass--dimension relation $m_\ph^2=\D(\D-2)$, the dual scalar therefore saturates the BF bound, $m_\ph^2=-1$ \cite{Breitenlohner:1982jf,Klebanov:1999tb}. At this bound, the two ultraviolet falloffs are degenerate, which leads to logarithmic corrections \cite{Bianchi:2001kw,Skenderis:2002wp}. The leading behaviors in the UV limit become  
\be
f(r)&=&\fr rR+\fr Rr\left[C_2\log^2\left(\fr rR\right)+C_1\log\left(\fr rR\right)+C_0\right]+\cdots,\nn
g(r)&=&\fr rR+\fr{BR}{r}+\cdots,\nn
\ph(r)&=&\fr{R}{r}\left[\b+\a\log\left(\fr rR\right)\right]+\cdots .
\la{Expansion: BFmetric}
\ee
Here $B$ is the normalizable coefficient of the temporal metric, and $C_0$, $C_1$, and $C_2$ describe the radial backreaction. In standard BF-bound quantization, $\alpha$ is the logarithmic source coefficient and $\beta$ is the finite coefficient corresponding to the vev of the operator. In general, their microscopic normalization depends on the scheme. However, we do not need it for the stress test, as will be seen later. The Einstein equations give
\be
C_2=\fr{\a^2}{4},\quad C_1=-\fr{\a^2}{4}+\fr{\a\b}{2},\quad {\rm and}\quad C_0=B+\fr{\a^2}{8}-\fr{\a\b}{4}+\fr{\b^2}{4}. \la{Result: BFmetriccoefficients}
\ee
The logarithmic terms produce a characteristic $\ell^2\log^2(R/\ell)$ correction to the short-distance entropy, derived in Appendix~\ref{Appendix: BFentropy}. This is an ultraviolet prediction of the Einstein-scalar description with a dimension-one operator, and we do not fit it to the finite-chain data.

We now construct an effective gravity for the off-critical TFIM from its entanglement data. The ultraviolet structure above restricts the asymptotic form of this description. However, it does not determine the geometry at a finite radius. The off-critical entanglement entropy gives this missing information. It fixes the spatial function $f(r)$ through the inverse Abel transformation. Once we know $f(r)$, the remaining bulk equations determine how this spatial geometry is completed into an effective spacetime. In particular, the temporal metric follows from the first integral of the Einstein equation. With $u_g(r)=g(r)/r$, we have
\be
u_g'(r)=\fr{C_g}{r^2f(r)}.
\la{Equation: temporalconstant}
\ee
The boundary time normalization fixes $g/r\to1/R$. For the ultraviolet expansion
\be
g(r)=\fr rR+\fr{BR}{r}+\cdots,
\ee
the integration constant is
\be
C_g=-2B.
\la{Result: CgB}
\ee
At this stage, $C_g$ is still a bulk integration constant. This point is important. Equal-time entanglement fixes the spatial geometry, but it does not determine the normalization of the temporal component. Therefore, the completion needs the bulk equations together with a physical infrared condition. The periodic TFIM entropy is symmetric under $\ell\to L-\ell$ and reaches its maximum at $\ell=L/2$, as shown in \Figref{Figure: TFIMentropy}. Within the continuous, smooth geodesic prescription, \Eqref{Equation: HamiltonJacobi} then gives $r_t\to0$ as $\ell\to L/2$. This motivates a completion that extends to $r=0$. A smooth, horizonless center is an additional physical condition on that completion, which entropy symmetry alone does not establish. We impose this condition in the effective Einstein-scalar model. In the areal gauge, it requires $f(0)=1$, $f'(0)=0$, $g(0)>0$, and $g'(0)=0$. Smooth rotationally invariant fields are even near the center, so we write
\be
\log f(r)=b_2r^2+b_4r^4+b_6r^6+\cdots.
\la{Expansion: regularf}
\ee
Integrating \Eqref{Equation: temporalconstant}, we obtain  
\be
g(r)=Kr-C_g[1+r{\cal H}(r)],
\la{Solution: regularGeneralG}
\ee
where $K$ is an integration constant and ${\cal H}(r)$ is defined as
\be
{\cal H}(r)=\int_0^r ds\,\fr{1-f(s)^{-1}}{s^2}.   \la{Relation: calH}
\ee
The odd term $Kr$ would violate $g'(0)=0$, so regularity requires $K=0$. The
ultraviolet normalization then determines
\be
C_g^{\rm reg}=-\fr{1}{R{\cal H}_\inf}\quad {\rm and}\quad
g_{\rm reg}(r)=\fr{1+r{\cal H}(r)}{R{\cal H}_\inf} ,
\la{Result: regularCg}
\ee
where ${\cal H}_\inf \equiv \lim_{r\to\infty}{\cal H}(r)$. This completes the gravity reconstruction. In summary, the entanglement entropy fixes the spatial metric, and the ultraviolet condition fixes the boundary normalization of time. Infrared regularity then selects the remaining integration constant.

Once we know $f$ and $g_{\rm reg}$, \Eqref{Equation: phi} determines $\ph'(r)^2$. A real canonical scalar requires this quantity to be nonnegative throughout the trajectory. This is a separate admissibility condition on the reconstructed geometry, beyond entropy concavity and the turning-point limit. Near a regular center, $\ph'(r)^2=q_2r^2+\cdots$ with $q_2\geq0$, and hence $\ph'(0)=0$. When the admissibility condition holds, the scalar profile is given by
\be
\ph(r)=\ph_0\pm\int^r d\widetilde r\,
\left[\fr{2}{\widetilde r}\left(
\fr{g'(\widetilde r)}{g(\widetilde r)}
-\fr{f'(\widetilde r)}{f(\widetilde r)}
\right)\right]^{1/2},
\la{Equation: scalarTrajectory}
\ee
where the source convention chooses the sign and $\ph(\infty)=0$ fixes $\ph_0$. The potential along the reconstructed trajectory then follows algebraically from \Eqref{Equation: V}. We now illustrate this construction with a representative off-critical state. We take $J=1$, $h=1.01$, and $L=512$, so that
\be
R=\fr{L}{2\pi}=81.487,\quad
m_\ps=2(h-J)=0.02,\quad {\rm and}\quad
\r\equiv\fr rR .
\la{Definition: rho}
\ee
\begin{figure}[t]
    \begin{center}
        \vspace{-0.5cm}
        \hspace{-0.5cm}
        \begin{minipage}[t]{0.45\textwidth}
        \centering
        \includegraphics[angle=0,width=\textwidth]{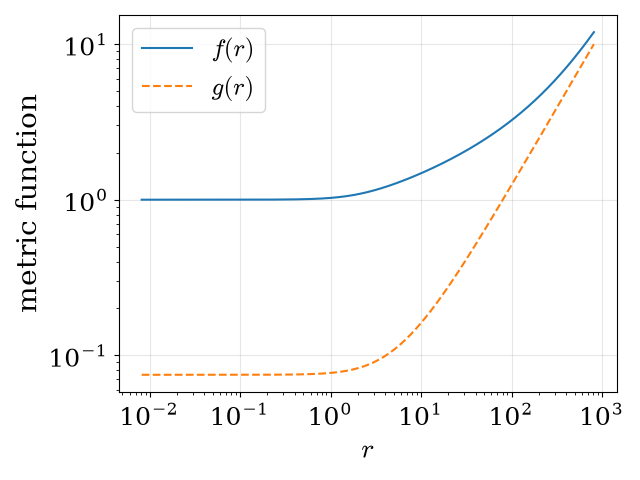}\\[-1mm]
        (a) $f(r) $ and $ g(r) $
        \end{minipage}
        \hspace{0.5cm}
        \begin{minipage}[t]{0.45\textwidth}
        \centering
        \includegraphics[angle=0,width=\textwidth]{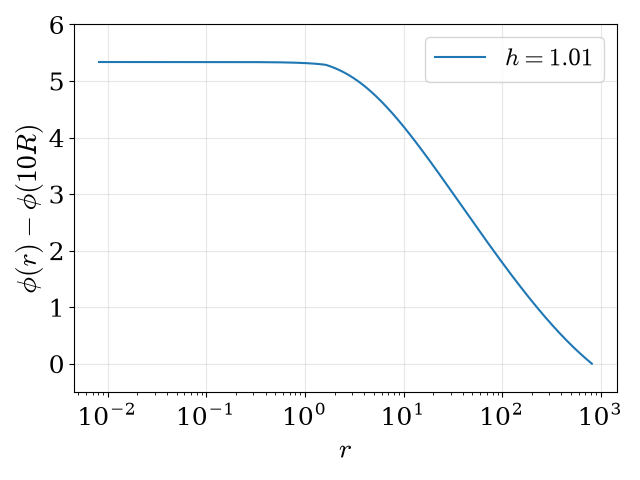}\\[-1mm]
        (b) $ \ph(r)-\ph(10 R) $
        \end{minipage}
        \vspace{-0cm}
            \caption{Representative off-critical reconstruction at $J=1$, $h=1.01$, and $L=512$. (a) Spatial metric $f$ from the entanglement entropy and temporal metric $g_{\rm reg}$ fixed by center regularity. (b) Scalar profile relative to its value at $r=10 R$.}
\la{Figure: representativeGeometry}
    \end{center}
\end{figure}

For this state, neither the entropy nor $f(r)$ has a simple closed form. We therefore evaluate the entropy from the covariance matrix and use the inverse Abel transformation to find the spatial metric with the same numerical prescription as in Sec.~\ref{Section: critical}. 
\Figsubref{Figure: representativeGeometry}{a} shows $f(r)$ together with $g_{\rm reg}(r)$ selected by the ultraviolet normalization and center regularity. \Figsubref{Figure: representativeGeometry}{b} shows the scalar profile that supports this geometry. With this symmetry-preserving inversion, the scalar kinetic term is positive on the investigated radial grid. When the half-chain entropy becomes nearly flat, the sign of $\ph'^2$ very close to the center becomes sensitive to the numerical treatment. This affects only the admissibility check and not the stress, because $C_g^{\rm reg}$ in \Eqref{Result: regularCg} depends on $f(r)$ alone.

\Figsubref{Figure: representativeEntropy}{a} shows the potential along the reconstructed trajectory, which follows from \Eqref{Equation: V}. Here we plot the potential against $\ph$ in the ultraviolet convention $\ph(\infty)=0$. Therefore, its scalar coordinate differs by a constant from the shifted profile $\ph(r)-\ph(10R)$ in \Figsubref{Figure: representativeGeometry}{b}. For a numerical consistency check, we confirm that the reconstructed spatial geometry reproduces the input entanglement entropy in \Figsubref{Figure: representativeEntropy}{b}. To do so, we compute the equal-time geodesics on the reconstructed geometry. 
Writing $\r=\r_t\cosh w$, the forward geodesic integrals are rewritten as
\be
\D\th(\r_t)
&=&2\int_0^{w_\L}\fr{dw}{\cosh w\,f(\r_t\cosh w)},\nn
S_{\rm RT}(\r_t)
&=&\fr c3\int_0^{w_\L}dw\,
\fr{\r_t\cosh w}{f(\r_t\cosh w)}\quad {\rm with}\quad
w_\L=\operatorname{arcosh}\left(\fr{\r_\L}{\r_t}\right).
\la{Equation: representativeForwardRT}
\ee
Here the finite radial cutoff changes both the additive entropy constant and the interval angle, so we keep the angle correction for the comparison. After fixing one additive constant, the two entropies agree to better than $3.5\times10^{-4}$ for intervals of at least ten sites. This checks the entropy values themselves. However, it does not provide an error bound on the half-chain second derivative or on the stress extracted from the geometry. \Figref{Figure: representativeGeometry} and \Figref{Figure: representativeEntropy} show that the reconstructed geometry is a regular solution with a real scalar, so the derivation of the stress in Sec.~\ref{Section: BrownYork} applies to it.

\begin{figure}[t]
    \begin{center}
        \vspace{-0.5cm}
        \hspace{-0.5cm}
        \begin{minipage}[t]{0.45\textwidth}
        \centering
        \includegraphics[angle=0,width=\textwidth]{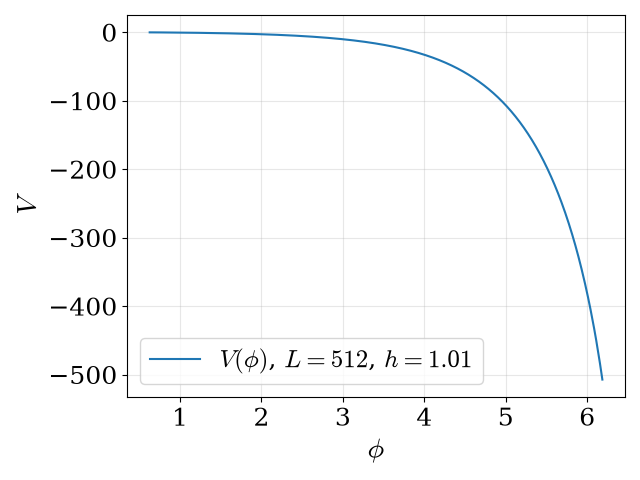}\\[-1mm]
        (a) $ V(\ph) $ 
        \end{minipage}
        \hspace{0.5cm}
        \begin{minipage}[t]{0.45\textwidth}
        \centering
        \includegraphics[angle=0,width=\textwidth]{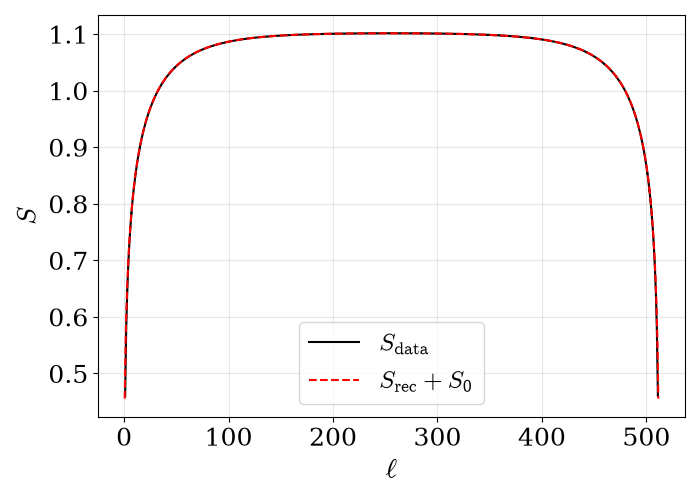}\\[-1mm]
        (b) $  S_E (n) $
        \end{minipage}
        \vspace{-0cm}
        \caption{Matter sector and entropy check for the representative geometry. (a) Scalar potential. (b) Exact TFIM entropy and forward-geodesic entropy.}
\la{Figure: representativeEntropy}
    \end{center}
\end{figure}

\subsection{Holographic enthalpy density and the stress test}\la{Section: BrownYork}

We now compare the stress predicted from the reconstructed dual gravity with the finite-size TFIM stress, which we calculate independently from the exact spectrum in Appendix~\ref{Appendix: TFIMpressure}. We did not use the microscopic stress to select the reconstructed geometry. On the gravity side, the normalizable part of the temporal metric encodes this observable. To read off this observable, we relate the temporal integration constant $C_g$ to the boundary stress tensor. We use the Brown-York tensor together with holographic
renormalization
\cite{Brown:1992br,Balasubramanian:1999re,deHaro:2000vlm,Bianchi:2001kw,Skenderis:2002wp}.
Related applications in Einstein-scalar and nonconformal holographic systems
can be found in
Refs.~\cite{Park:2018ebm,Park:2024fik}.
We take a timelike regulating surface $\S_r$ at large $r$. Its induced metric
and outward unit normal are
\be
\g_{ab}dx^a dx^b=-g^2dt^2+r^2d\th^2\quad {\rm and}\quad n^M\pa_M=f\,\pa_r .
\la{Result: BYinducedmetric}
\ee
The extrinsic curvature of $\S_r$ is $K_{ab}=e_a{}^M e_b{}^N\nabla_M n_N$ with
the tangent vectors $e_a{}^M=\pa X^M/\pa x^a$. For our metric ansatz, it becomes
\be
K_{tt}=-fgg'\quad {\rm and}\quad K_{\th\th}=fr\quad {\rm with}\quad K=f\left(\fr{g'}g+\fr1r\right).
\la{Result: BYextrinsiccurvature}
\ee
With the outward normal in \Eqref{Result: BYinducedmetric}, the
counterterm-renormalized Brown-York tensor on $\S_r$ is
\be
T_{ab}^{\rm UV}
=\fr1{8\pi G}
\left(K\g_{ab}-K_{ab}-\fr1R\g_{ab}\right)
+T_{ab}^{(\ph,{\rm ct})} ,
\la{Result: BYtensor}
\ee
where the scalar term contains the logarithmic counterterms at the BF bound
and finite local covariant counterterms
\cite{deHaro:2000vlm,Bianchi:2001kw,Skenderis:2002wp,Papadimitriou:2004rz}.

To read off the boundary stress, we separate the logarithmic BF-bound terms in
the ultraviolet expansion in \Eqref{Expansion: BFmetric}. We write the
remaining finite parts as
\be
g=\fr rR+B\fr Rr+\cdots\quad {\rm and}\quad
f=\fr rR+C_0\fr Rr+\cdots ,
\la{Result: BYfiniteUVcoefficients}
\ee
where $C_0$ is the finite nonlogarithmic coefficient of the radial
metric function in \Eqref{Expansion: BFmetric}. We multiply the orthonormal components of
\Eqref{Result: BYtensor} by $(r/R)^2$ and take $r\to\inf$. Then the boundary
energy density and pressure are
\be
\e=-\fr{C_0}{8\pi GR}+\e_{\rm loc}\quad {\rm and}\quad
p=\fr{C_0-2B}{8\pi GR}+p_{\rm loc} ,
\la{Result: BYstresscomponents}
\ee
where $\e_{\rm loc}$ and $p_{\rm loc}$ come from the local scalar counterterms and depend on the scheme. For the homogeneous source considered here, derivative counterterms vanish, and the remaining counterterms are proportional to $\g_{ab}$. This implies $\e_{\rm loc}+p_{\rm loc}=0$. Therefore, both $C_0$ and the scheme-dependent terms cancel in the enthalpy density \cite{Bianchi:2001kw,Skenderis:2002wp,Papadimitriou:2004rz}. Together with $C_g=-2B$ in \Eqref{Result: CgB}, we finally obtain
\be
C_g=-2B=8\pi GR(\e+p).
\la{Result: BrownYorkTemporalNormalization}
\ee
For global AdS, $C_0=B=1/2$, so $\e=p=-1/(16\pi GR)$ and $C_g=-1$.
This agrees with the Casimir energy of the critical Ising theory in Sec.~\ref{Section: critical},
and it fixes the sign and the normalization of the stress test below. Since $C_g$ gives the enthalpy density through \Eqref{Result: BrownYorkTemporalNormalization}, which does not depend on the scheme, we call $C_g$ the stress coefficient.

In Sec.~\ref{Section: fullGeometry}, we determined the spatial geometry from the entanglement entropy data and its temporal completion, requiring regularity at the center. Then, the gravitational stress follows without using the microscopic stress. Using holographic renormalization, we obtain the gravitational enthalpy density
\be
(\e+p)_{\rm grav}=\fr{C_g^{\rm reg}}{8\pi GR},
\la{Result: gravStressFromCg}
\ee
where, from \Eqref{Result: regularCg}, the gravitational stress coefficient is
\be
C_g^{\rm reg} =-\fr{1}{R {\cal H}_\infty(f_{S_E})}.
\la{Result: gravCgReg}
\ee
This prediction rests on the following conditions, which we use throughout the rest of the paper. The solution is a static, rotationally invariant solution of the minimal canonical Einstein-scalar model, with a real scalar, a smooth horizonless center, and the ultraviolet time normalization $g/r\to1/R$, and the equal-time entropy is assigned by the continuous geodesic prescription. Under these conditions, \Eqref{Result: gravCgReg} fixes the holographic stress from the spatial geometry. For the pure state on the boundary circle, $S_E(\ell)=S_E(L-\ell)$ and $S_E'(L/2)=0$, while strong subadditivity gives $S_E''(L/2)\le0$ \cite{Casini:2004bw}. We now express the stress entirely through this entropy, using the geodesic and Abel framework \cite{Bilson:2008ab,Bilson:2010ff,Huh:2026nvt}. 
Subtracting the flat-center contribution from \Eqref{Relation: turningAngle} gives
\be
\D\th(r_t)=\pi-2r_t\int_{r_t}^{\inf}dr\, \fr{ 1 - f(r)^{-1} }{r \sqrt{r^2-r_t^2}}.
\la{Equation: centerAngleSubtraction}
\ee
In the $r_t\to0$ limit, the integral approaches ${\cal H}_\inf$, because $f(r)=1+{\cal O}(r^2)$ near the center and $f(r)\sim r/R$ in the ultraviolet. Therefore, $\D\th$ is rewritten as
\be
\D\th(r_t)=\pi-2r_t{\cal H}_\inf+\cdots .
\la{Equation: centerAngleLimit}
\ee
Combining this result with the Hamilton-Jacobi relation in \Eqref{Equation: HamiltonJacobi}, we obtain the following relation
\be
\left.\fr{d^2S_E}{d\ell^2}\right|_{\ell=L/2}
=-\fr{1}{8GR^2{\cal H}_\inf} ,
\la{Equation: halfIntervalCurvature}
\ee
which with Eqs.~\eq{Result: gravStressFromCg} and \eq{Result: gravCgReg} results in
\be
(\e+p)_{\rm grav}
=\fr1\pi\left.\fr{d^2S_E}{d\ell^2}\right|_{\ell=L/2} ,
\la{Result: curvatureStress}
\ee
which shows that the entropy curvature at half the system size fixes the holographic enthalpy density. 

\Eqref{Result: curvatureStress} is not a new input. It is a rewriting of the holographic enthalpy density in \Eqref{Result: gravStressFromCg} in terms of the boundary entropy. At the critical point, \Eqref{Result: curvatureStress} reduces to the CFT result in \Eqref{Result: criticalCurvatureStress}. On the dual gravity side, a different potential changes $f(r)$ and thus both sides of \Eqref{Result: curvatureStress}, but the relation holds for every solution that satisfies the conditions above. The potential therefore enters implicitly, through $f(r)$ and through the BF-bound counterterms that make $\epsilon+p$ independent of the scheme. In this work, the entanglement entropy fixes $f(r)$ directly. Note that the Newton constant cancels in \Eqref{Result: curvatureStress}, so the relation does not depend on the normalization $3R/(2G)=1/2$, although $c=1/2$ lies far from the large-$c$ regime where classical gravity is usually justified. Therefore, \Eqref{Result: curvatureStress} is a prediction of the effective holographic description, which can be tested directly against the microscopic TFIM.

\Figref{Figure: stressDependence} presents the central result of this work, the comparison of the stress predicted by the entanglement-based geometry with the exact TFIM stress, which is not used in the reconstruction. We plot $2B=-C_g$, calculated from the reconstructed geometry, and use \Eqref{Result: curvatureStress} as a closed-form check. \Figsubref{Figure: stressDependence}{a} compares the holographic prediction with the exact TFIM stress for $L=128$, $256$, and $512$, while \Figsubref{Figure: stressDependence}{b} displays the same data against $x=m_\ps L/v$. The three sizes follow a common scaling curve, and the gravity prediction agrees closely with the exact stress in the sampled near-critical scaling regime. For example, at $L=512$ and $h=1.01$, the reconstructed geometry gives $2B=0.0747896$, while the exact spectrum gives $2B=0.0747898$. The remaining small difference is a lattice correction, which decreases as $N^{-2}$ at fixed $x$, as shown in Appendix~\ref{Section: latticeCorrection}. Therefore, \Eqref{Result: curvatureStress}, which is predicted by the effective holographic description, holds within numerical accuracy for the free massive Majorana theory, the continuum limit of the TFIM.

\begin{figure}[t]
    \begin{center}
        \vspace{-0.35cm}
        \hspace{-0.35cm}
        \begin{minipage}[t]{0.45\textwidth}
        \centering
        \includegraphics[angle=0,width=\textwidth]{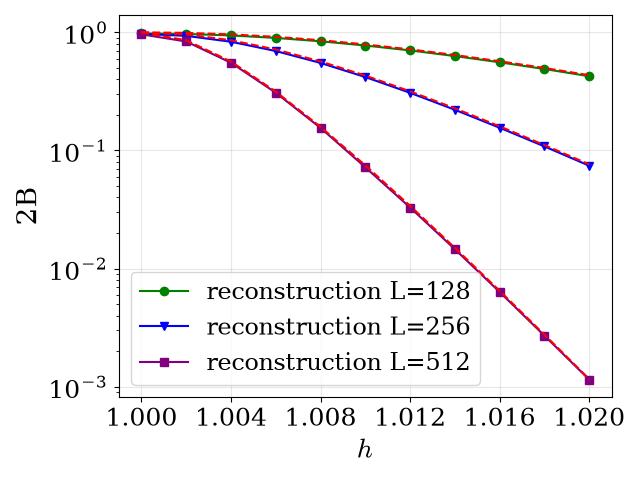}\\[0.8mm]
        \textcolor{black}{(a) $2B(h)$}
        \end{minipage}
        \hspace{0.55cm}
        \begin{minipage}[t]{0.45\textwidth}
        \centering
        \includegraphics[angle=0,width=\textwidth]{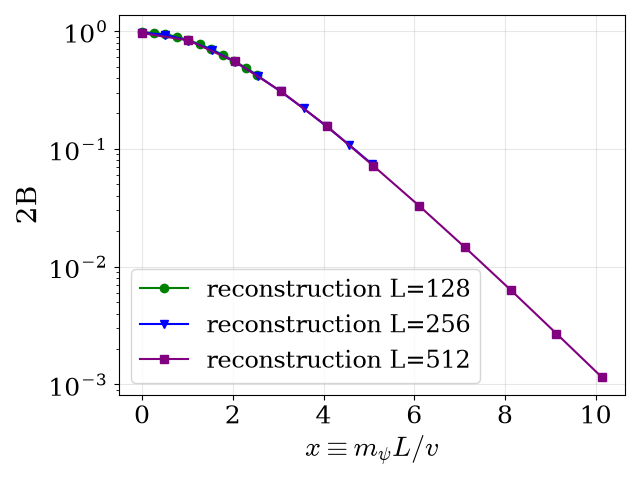}\\[0.8mm]
        \textcolor{black}{(b) Scaling in $m_\ps L/v$}
        \end{minipage}
        \vspace{0.1cm}
        \hspace{-0.35cm}
                \vspace{0.02cm}
        \caption{Independent stress test of the entanglement-based effective geometry, the central result of this work. Symbols show the holographic prediction $2B=1/(R{\cal H}_\inf)$, and red dashed curves show the exact TFIM stress, which is not used in the reconstruction. (a) Dependence on $h$ for $L=128$, $256$, and $512$. (b) The same data against $x=m_\ps L/v$, showing their common scaling behavior.}\la{Figure: stressDependence}
    \end{center}
\end{figure}

\subsection{Universal scalar potential of the reconstructed gravity}
\la{Section: universalPotential}

We now show that, in the near-critical scaling regime, the reconstructed dual gravity is described by one universal scalar potential. Each reconstruction uses the entanglement entropy of one state, specified by $h$ and $L$, and gives the potential only along its own trajectory, as in \Figsubref{Figure: representativeEntropy}{a}. If the reconstructed geometries are solutions of one effective Einstein-scalar theory, the potentials obtained from different states must lie on one curve, and $h$ and $L$ change only the solution, for example how deep the trajectory goes into the potential. The stress test in Sec.~\ref{Section: BrownYork} does not check this, because the stress depends only on $f(r)$. Since the finite-size physics in the scaling regime depends on $h$ and $L$ only through $x=m_\ps L/v$, as shown for the stress in \Figsubref{Figure: stressDependence}{b} and for the reconstructed trajectory below, it is enough to compare reconstructions with different $x$. We reconstruct four geometries at $J=1$ and $h=1.004$ with $L=512$, $768$, $1024$, and $1536$, which cover $2\lesssim x\lesssim6$. All of them use the canonical kinetic normalization and the same sign of the scalar, but the numerical reconstruction does not fix a common ultraviolet origin of the scalar. To compare only the shape of the potential, we shift the scalar of each trajectory as
\be
\widehat\ph=\ph-\ph_*\quad {\rm with}\quad V(\ph_*)=-2 ,
\la{Definition: potentialAlignment}
\ee
where $V=-2$ is chosen inside the range in which all compared curves are monotonic and overlap. This shift removes only the additive ambiguity of the scalar, and we neither rescale the scalar nor shift the potential vertically. Since it differs from the ultraviolet convention $\ph(\infty)=0$ of \Figsubref{Figure: representativeEntropy}{a}, it does not test the source normalization at the BF bound.

After the alignment, the four curves overlap on $-1\lesssim\widehat\ph\lesssim0.78$, which lies between the outer end of the $L=1536$ curve and the inner end of the $L=512$ curve. As shown in \Figsubref{Figure: universalPotential}{a}, they almost coincide there and deviate slightly only near the inner end, where the $L=512$ curve bends before it ends. Taking the $L=1536$ curve as a reference, the largest difference of $\widehat\ph$ at a fixed $V$ is about $0.03$, $0.004$, and $0.002$ for $L=512$, $768$, and $1024$, respectively. Therefore, the four sizes share one common shape of the potential, and the deviation becomes smaller quickly as $L$ increases.

\begin{figure}[t]
    \begin{center}
        \vspace{-0.35cm}
        \hspace{-0.35cm}
        \begin{minipage}[t]{0.45\textwidth}
        \centering
        \includegraphics[angle=0,width=\textwidth]{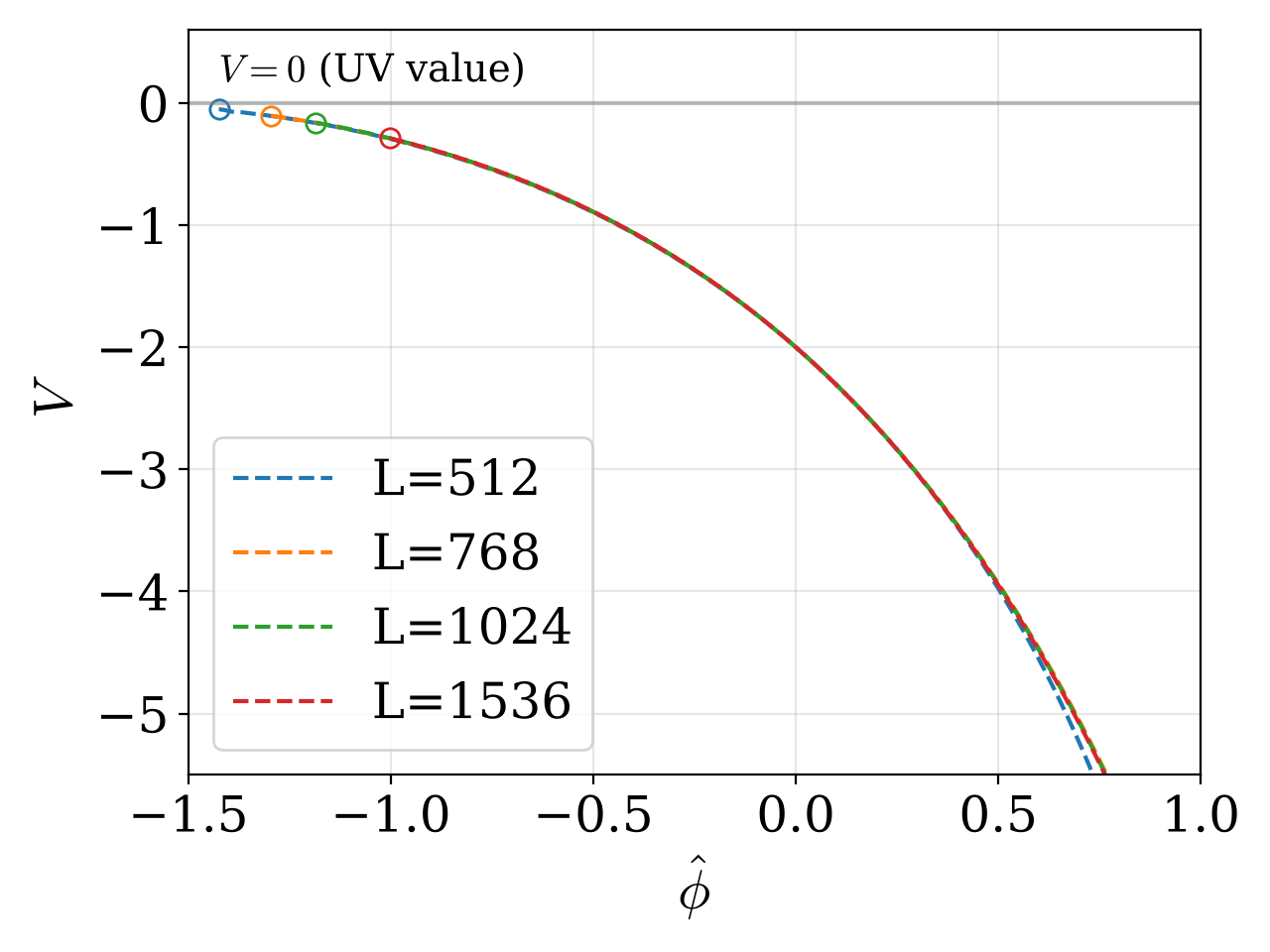}\\[0.8mm]
        \textcolor{black}{(a) Common potential}
        \end{minipage}
        \hspace{0.55cm}
        \begin{minipage}[t]{0.45\textwidth}
        \centering
        \includegraphics[angle=0,width=\textwidth]{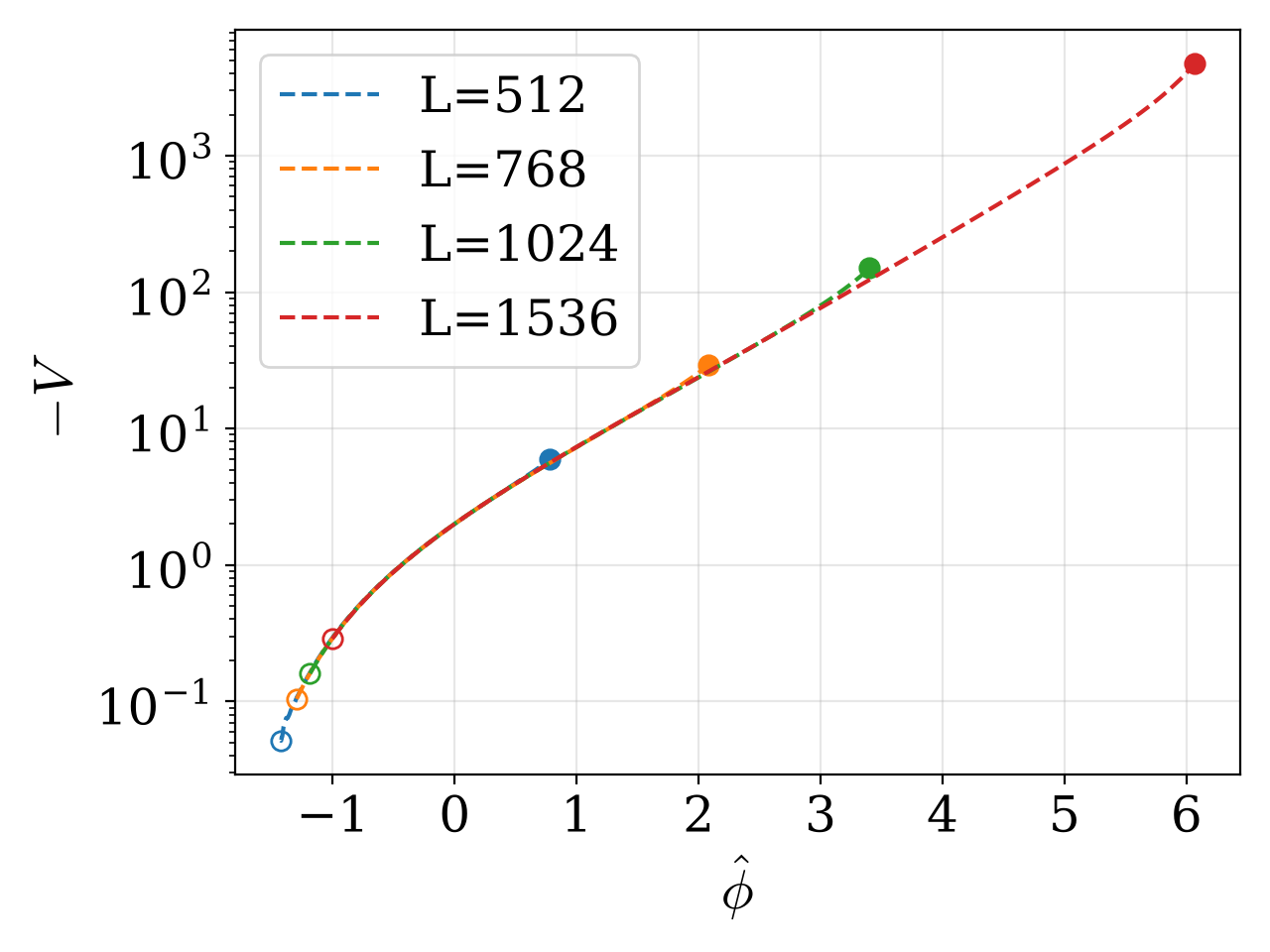}\\[0.8mm]
        \textcolor{black}{(b) Full trajectories}
        \end{minipage}
        \vspace{0.1cm}
        \hspace{-0.35cm}
                \vspace{0.02cm}
        \caption{Common scalar potential at $J=1$ and $h=1.004$. (a) Curves aligned by \Eqref{Definition: potentialAlignment}. (b) Full trajectories on a logarithmic scale. Filled dots mark the inner end of each trajectory, and open circles mark the outer end $\r=20$.}\la{Figure: universalPotential}
    \end{center}
\end{figure}

\Figsubref{Figure: universalPotential}{b} shows the full trajectories on a logarithmic scale. Near the center, each finite-size curve bends away from the common curve before it ends, and we call this bent part the hook. As $L$ increases, the end of the hook moves from $\widehat\ph\simeq0.78$ at $L=512$ to $\widehat\ph\simeq6.07$ at $L=1536$, while the common part of the curves does not move, which supports a finite-size origin of the hook. The whole trajectory, including the hook, also depends on $h$ and $L$ only through $x$. For example, $(h,L)=(1.004,1024)$ and $(1.008,512)$ have almost the same $x$. Their aligned potentials differ by less than $2\times10^{-4}$ in $\widehat\ph$ for $V\le-1$, and their hooks end at almost the same position, $\widehat\ph\simeq3.40$ and $3.39$. Therefore, changing $L$ at fixed $h$ is equivalent to changing $h$ at fixed $L$, and the size dependence in \Figref{Figure: universalPotential} also represents the coupling dependence over the same range of $x$. As a result, a single universal potential describes the reconstructions for different sizes and couplings in the sampled scaling regime, as expected for an effective theory in which the action is fixed and $h$ and $L$ enter only through the solution.

\section{Discussion}\la{Section: discussion}

We have studied how far the equal-time entanglement of a known microscopic Hamiltonian can determine a predictive effective gravity. The Ryu-Takayanagi relation determines the spatial metric from equal-time entanglement. The Einstein-scalar equations and physical boundary conditions then select the temporal geometry. 
The construction is not tied to the TFIM. Within the static symmetry class considered here, the same logic applies when sufficiently resolved entanglement data are available, and the reconstructed spatial geometry admits a consistent gravitational completion.

The critical and off-critical calculations play different roles. At the Ising critical point, the exact entropy reconstructs global AdS spatial geometry and checks the stress normalization. Away from criticality, the stress comparison tests a prediction of the reconstructed effective geometry. The dimension $\D=1$ of the Ising energy-density operator places the bulk scalar at the AdS$_3$ BF bound. Its degenerate ultraviolet modes give logarithmic backreaction and the short-distance $\ell^2\log^2(R/\ell)$ entropy correction in Appendix~\ref{Appendix: BFentropy}. This analytic ultraviolet prediction is independent of the off-critical stress test, and it agrees with the squared-logarithm term of the massive Majorana theory, which fixes the source coefficient by the fermion mass.

The central result is the independent off-critical stress comparison in \Figref{Figure: stressDependence}. The entanglement-based geometry closely reproduces the coupling dependence and finite-size scaling of the exact TFIM stress in the sampled near-critical regime, even though the microscopic stress is not an input. Under the conditions of Sec.~\ref{Section: BrownYork}, \Eqref{Result: curvatureStress} expresses this geometric prediction as an entanglement-to-stress relation involving only the half-system entropy curvature. The status of this relation is therefore clear. It is predicted by the effective holographic description, and it holds within numerical accuracy for the free massive Majorana theory, because the remaining finite-lattice discrepancy decreases as $N^{-2}$ toward the continuum limit (Appendix~\ref{Section: latticeCorrection}). A direct field-theory derivation remains open.

The discussion of the universal potential in Sec.~\ref{Section: universalPotential} adds a second test, which is complementary to the stress test. The stress test checks the part of the construction that does not depend on the potential. On the other hand, the universal potential checks the part that does. The reconstructions for different sizes and couplings share one potential, and $h$ and $L$ enter only through the solution, which depends on $x$. The only deviation is the hook near the center, which moves away as $x$ increases, as expected for a finite-size effect. Therefore, the reconstructed geometries behave as solutions of one effective Einstein-scalar theory, rather than as independent fits to each set of entanglement data.

The numerical test covers the disordered branch $h>J$, whose scaling limit is a free massive Majorana theory.  A field-theory derivation and tests in interacting spin chains would clarify the range of validity and whether the relation extends beyond free fermions. The stress agreement and the universal potential do not establish a classical gravity dual of the TFIM. However, they show that an effective geometric construction can behave as one consistent effective theory and generate a relation between field-theory observables that is independently testable.

\vspace{0.5cm}

\vspace{1.5cm}

{\bf Acknowledgments}

CP was supported by the National Research Foundation of Korea (NRF) grant funded by the Korea government (MSIT) (No. RS-2026-25473667). DHK and JY acknowledge the support from the National Research Foundation of Korea (Grant No. RS-2024-00392445), funded by the Korea government (MSIT). This work is also partially supported by the ANCHOR program through the Gwangju ANCHOR Center, funded by the Ministry of Education (MOE) and the Jeonnam-Gwangju Special Metropolitan City, Republic of Korea (2026-ANCHOR-05-001).

\vspace{0.5cm}

\appendix

\section{Short-interval entanglement entropy at the BF bound}
\la{Appendix: BFentropy}

This appendix derives the short-distance correction to the entanglement entropy produced by the logarithmic backreaction at the BF bound. The equal-time spatial part of the off-critical metric in \Eqref{Ansatz: metric} has the same form as the critical spatial metric in \Eqref{Ansatz: criticalSpatialMetric}. Therefore, the holographic entropy and the turning-point relations in Eqs.~\eq{Equation: HoloEn}-\eq{Relation: turningEntropy} remain valid, with $f(r)$ now given by the off-critical radial function. For a short interval, the geodesic stays in the asymptotic region. Thus we can evaluate the entropy with the ultraviolet expansion of the radial metric in \Eqref{Expansion: BFmetric}, which gives
\be
\fr{1}{f(r)^2}=\fr{R^2}{r^2}
\left[1-\fr{2R^2}{r^2}P(r)+\cdots\right]  \quad {\rm with} \quad
P(r)=C_2\log^2\left(\fr rR\right)
+C_1\log\left(\fr rR\right)+C_0 .
\ee
As a result, in the short-interval limit, the entanglement entropy expands as
\be
S_E(\ell)=\fr c3\log\left(\fr{\ell}{a}\right)
-\fr{c\ell^2}{36R^2}{\cal F}_{\rm UV}
+{\cal O}\left(\fr{\ell^4}{R^4}\log^4\fr R\ell\right),
\la{Result: UVentropy}
\ee
where
\be
{\cal L}_\ell &=&\log\left(\fr R\ell\right)+\fr43  \ ,  \quad 
{\cal C}=\fr{10}{9}-\fr{\pi^2}{12}, \nn
{\rm and}\quad {\cal F}_{\rm UV}&=&B+\fr{\b^2}{4}-\a\b\left(\fr14-\fr{{\cal L}_\ell}{2}\right)+\a^2\left[\fr14\left({\cal L}_\ell^2+{\cal C}\right)-\fr{{\cal L}_\ell}{4}+\fr18\right].
\la{Result: UVfunction}
\ee
Here the first term in \Eqref{Result: UVentropy} is the leading contribution of the Ising CFT. The term quadratic in the BF source gives the first correction,
\be
\d S_E=-\fr{c\a^2}{144}
\fr{\ell^2}{R^2}\log^2\left(\fr R\ell\right).
\la{Result: UVleadinglog}
\ee
Thus the $\ell^2\log^2(R/\ell)$ correction follows from the BF-bound scalar falloffs in the effective Einstein-scalar model. It connects the dimension-one deformation to the ultraviolet structure of that description, and we compare it with the massive Majorana theory below. It is consistent with earlier analyses of logarithmic holographic corrections and relevant perturbations \cite{Park:2021nyc,Park:2015dia,Nishioka:2014kpa,Herzog:2013py}. As a simple check, setting $\a=\b=0$ and $B=1/2$ gives global AdS, and the entropy reduces to
\be
S_E= \frac{c}{3} \log \ls \frac{\ell}{a} \rs - \frac{c \pi^2 \ell^2}{18 L^2} +\cdots , 
\ee
where the second term is a finite-size correction. This agrees with the
short-interval expansion of \Eqref{Result: CFTEn}.

The massive Majorana theory gives the leading short-distance correction \cite{Casini:2005rm}
\be
\d S_E=-\fr{c}{6}\left(\fr{m_\ps\ell}{v}\right)^2\log^2\left(\fr{m_\ps\ell}{v}\right)+\cdots \quad{\rm with}\quad c=\fr12 .
\la{Result: MajoranaLog}
\ee
Its squared-logarithm term agrees with the BF-bound form in \Eqref{Result: UVleadinglog}. Matching the coefficients fixes the bulk source normalization
\be
\a^2=24\left(\fr{m_\ps R}{v}\right)^2\quad{\rm or}\quad |\a|=\fr{\sqrt6}{\pi}\,x ,
\la{Result: alphaMass}
\ee
where $x=m_\ps L/v$. Thus the short-distance entropy relates the bulk source to the fermion mass. This consistency check is separate from the stress test, which does not use $\a$.

\section{Exact TFIM spectrum and finite-size stress}
\la{Appendix: TFIMpressure}

For periodic spin boundary conditions, the Jordan-Wigner solution separates the finite chain into fermion-parity sectors. We use the disordered branch $h>J$ throughout the numerical stress test. In this case, the physical ground state becomes the even-parity Neveu-Schwarz quasiparticle vacuum with the following allowed momenta
\be
\bar k_n^{\rm NS}=\fr{(2n+1)\pi}{N}.
\la{Appendix: NSmomenta}
\ee
The exact single-particle dispersion is
\be
\o(\bar k)=2\sqrt{J^2+h^2-2Jh\cos\bar k},
\la{Appendix: exactdispersion}
\ee
and the ground-state lattice energy used in the calculation is
\be
E_0^{\rm NS}(N,J,h)
=-\fr12\sum_{n=0}^{N-1}\o(\bar k_n^{\rm NS})
=-\sum_{n=0}^{N-1}\sqrt{J^2+h^2-2Jh\cos\fr{(2n+1)\pi}{N}}.
\la{Appendix: exactgroundenergy}
\ee
We do not need the Ramond sector for the data on the disordered branch. Its competition with the Neveu-Schwarz sector becomes important on the ordered side, which lies outside the present numerical test.

We next isolate the finite-size contribution. The same dispersion gives the thermodynamic energy per site
\be
e_\inf(J,h)=-\fr{1}{2\pi}\int_{-\pi}^{\pi}d\bar k\,\sqrt{J^2+h^2-2Jh\cos\bar k}=-\fr{2(J+h)}{\pi}\,{\bf E}\left(\fr{4Jh}{(J+h)^2}\right), \la{Appendix: bulkenergy}
\ee
where ${\bf E}(m)$ is the complete elliptic integral
\be
{\bf E}(m)=\int_0^{\pi/2}d\vartheta\,\sqrt{1-m\sin^2\vartheta}.
\ee
We define the finite-size contribution by
\be
E_{\rm Cas}(N,J,h)
=E_0^{\rm NS}(N,J,h)-N e_\inf(J,h).
\la{Appendix: Casimirenergy}
\ee
This subtraction removes the microscopic extensive vacuum energy without changing the universal finite-size term.

We now use the finite-size energy to determine the boundary stress. The relativistic boundary time is $\ta=v t_{\rm lat}$ with $v=2\sqrt{Jh}\,\bar a$. Therefore, the energy conjugate to $\ta$ is $E_{\rm rel}=E_{\rm Cas}/v$. At fixed $h$, $J$, and $\bar a$, the renormalized energy density and pressure are
\be
\e=\fr{E_{\rm Cas}}{vL}\quad {\rm and}\quad
p=-\fr1v
\left.\fr{\pa E_{\rm Cas}}{\pa L}
\right|_{h,J,\bar a}.
\la{Appendix: latticepressure}
\ee
The derivative is a mechanical size derivative at fixed Hamiltonian
couplings. Combining \Eqref{Appendix: latticepressure} with $c=3R/(2G)$ and
$L=2\pi R$ gives
\be
C_g^{\rm TFIM}
=8\pi GR(\e+p)
=\fr{3}{\pi cv}
\left[
LE_{\rm Cas}
-L^2\fr{\pa E_{\rm Cas}}{\pa L}
\right]_{h,J,\bar a}.
\la{Appendix: exactCgpressure}
\ee

\section{Lattice correction and the continuum limit}\la{Section: latticeCorrection}

\begin{figure}[t]
	\begin{center}
		\vspace{-0.5cm}
		\hspace{-0.5cm}
		\includegraphics[angle=0,width=0.5 \textwidth]{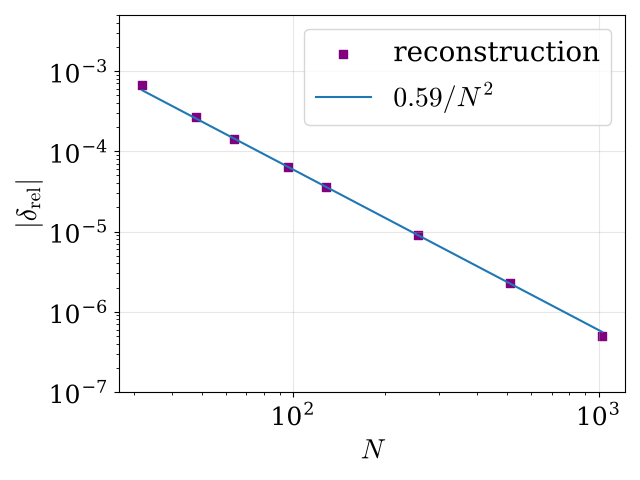}
		\vspace{-0cm}
		\caption{Continuum limit of the stress test at fixed $x=m_\ps L/v=3$. Symbols show $|\delta_{\rm rel}|$ obtained from the reconstructed geometry, and the line shows the $N^{-2}$ behavior.}
		\la{Figure: continuumLimit}
	\end{center}
\end{figure}

The relation in \Eqref{Result: curvatureStress} is a prediction for the continuum theory, which is the massive Majorana theory. However, the TFIM is defined on a lattice with spacing $\bar a$, so the exact TFIM stress also contains lattice corrections. Their size is set by the lattice spacing in units of the correlation length $\xi=v/m_\ps$,
\be
\fr{\bar a}{\xi}=\fr{m_\ps\bar a}{v}= \fr{h-J}{\sqrt{hJ}} ,
\ee
and the leading lattice corrections are of relative order $(\bar a/\xi)^2$. These corrections are different from the finite-size effect controlled by $L/\xi=m_\ps L/v=x$. The finite-size stress is a physical property of the continuum theory on a circle, and it is the quantity we test. On the other hand, the lattice corrections are artifacts of the lattice, and they must vanish in the continuum limit $\bar a/\xi\to0$.

To separate the two effects, we keep $x$ fixed and increase the number of sites $N$ with $L=N\bar a$. Then the continuum state does not change, and only $\bar a/\xi=x/N$ decreases. If the holographic relation holds in the continuum theory and the leading lattice correction dominates the reconstruction error, the discrepancy is expected to decrease as $N^{-2}$. We measure this difference by
\be
\delta_{\rm rel}(x,N)=
\frac{(2B)_{\rm grav}-(2B)_{\rm TFIM}}{(2B)_{\rm TFIM}},
\la{Result:relativeError}
\ee
where $(2B)_{\rm TFIM}=-C_g^{\rm TFIM}$ is the exact TFIM stress defined in Appendix~\ref{Appendix: TFIMpressure}. In \Figref{Figure: continuumLimit}, we evaluate $(2B)_{\rm grav}=1/(R{\cal H}_\inf)$ from the holographic enthalpy density of the reconstructed geometry. Evaluating the same prediction directly from the input entropy curvature with \Eqref{Result: curvatureStress} gives the same discrepancies to about one percent, as expected from \Eqref{Equation: halfIntervalCurvature}. At fixed $x=3$, the discrepancy follows $|\delta_{\rm rel}|\simeq0.59\,N^{-2}$, consistent with the expected $N^{-2}$ lattice correction. At the largest size $N=1024$, the discrepancy lies about ten percent below this line, which is likely due to the finite numerical precision of the entropy at this size. These results support convergence toward the continuum relation.


%

\end{document}